\documentclass[aps,prd,preprint,superscriptaddress,showpacs]{revtex4}
\usepackage{graphicx}

\usepackage{ulem}
\usepackage{color}
\definecolor{My_red}        {cmyk}{0.00,1.00,1.00,0.20}

\newcommand{\bmat}{\left(\begin{array}}
\newcommand{\emat}{\end{array}\right)}
\newcommand{\beq}{\begin{equation}}
\newcommand{\eeq}{\end{equation}}
\newcommand{\wt}{\widetilde}

\def\bwt{\begin{widetext}}
\def\ewt{\end{widetext}}
\def\be{\begin{equation}}
\def\ee{\end{equation}}
\def\bea{\begin{eqnarray}}
\def\eea{\end{eqnarray}}
\def\bean{\begin{eqnarray*}}
\def\eean{\end{eqnarray*}}
\def\bary{\begin{array}}
\def\eary{\end{array}}
\def\bit{\begin{itemize}}
\def\eit{\end{itemize}}

\def\lan{\langle}
\def\ran{\rangle}

\def\su5u1{SU(5) \times U(1)}
\def\fsu5u1{SU(5) \times U(1)'}
\def\so10{SO(10)}
\def\sq20{SO(10) \times SO(10)}

\def\bwt{\begin{widetext}}
\def\ewt{\end{widetext}}
\def\be{\begin{equation}}
\def\ee{\end{equation}}
\def\bea{\begin{eqnarray}}
\def\eea{\end{eqnarray}}
\def\bean{\begin{eqnarray*}}
\def\eean{\end{eqnarray*}}
\def\bary{\begin{array}}
\def\eary{\end{array}}
\def\bit{\begin{itemize}}
\def\eit{\end{itemize}}

\def\lan{\langle}
\def\ran{\rangle}

\def\su5u1{SU(5) \times U(1)}
\def\fsu5u1{SU(5) \times U(1)'}
\def\so10{SO(10)}
\def\sq20{SO(10) \times SO(10)}

\usepackage[centertags]{amsmath}
\usepackage{amssymb}

\begin{document}

\title{ Decaying Dark Matter in the Supersymmetric Standard Model
with Freeze-in and Seesaw mechanims}

\author{Zhaofeng Kang}

\affiliation{Key Laboratory of Frontiers in Theoretical Physics,
      Institute of Theoretical Physics, Chinese Academy of Sciences,
Beijing 100190, P. R. China }

\author{Tianjun Li}

\affiliation{Key Laboratory of Frontiers in Theoretical Physics,
      Institute of Theoretical Physics, Chinese Academy of Sciences,
Beijing 100190, P. R. China }

\affiliation{George P. and Cynthia W. Mitchell Institute for
Fundamental Physics, Texas A$\&$M University, College Station, TX
77843, USA }

\date{\today}

\begin{abstract}

Inspired by the decaying dark matter (DM) which can explain cosmic
ray anomalies naturally, we consider the supersymmetric Standard
Model with three right-handed neutrinos (RHNs) and $R$-parity, and
introduce a TeV-scale DM sector with two fields $\phi_{1,2}$ and
a $Z_3$  discrete symmetry. The DM sector only interacts with the RHNs
via a very heavy field exchange and then we can explain the cosmic ray
anomalies. With the second right-handed neutrino $N_2$ dominant
seesaw mechanism at the low scale around $10^4$ GeV,
we show that $\phi_{1,2}$ can obtain the vacuum expectation
values around the TeV scale, and then the lightest state from
$\phi_{1,2}$ is the decay DM with lifetime around $\sim 10^{26}s$.
In particular, the DM very long lifetime is related to the
tiny neutrino masses, and the dominant DM decay channels to
$\mu$ and $\tau$ are related to the approximate $\mu-\tau$ symmetry.
Furthermore, the correct DM relic density can be obtained via
the freeze-in mechanism, the small-scale problem for power
spectrum can be solved due to the decays of the $R$-parity
odd meta-stable states in the DM sector, and the baryon asymmetry
can be generated via the soft leptogensis.

\end{abstract}

\pacs{12.60.Jv, 14.70.Pw, 95.35.+d}

%\preprint{MIFP-10-nn}

\maketitle

\section{Introduction and Motivation}

The cosmic ray anomalies  observed by PAMELA and
Fermi-LAT~\cite{Adriani:2008zr,Abdo:2009zk} strongly indicated
that the dark matter (DM) particles annihilate or decay
dominantly into the leptons. Although the large annihilation
cross sections can be realized via the Sommerfield enhancement
or Breit-Wigner mechanism~\cite{ArkaniHamed:2008qn},
the HESS obervation~\cite{Collaboration:2008aaa, Aharonian:2009ah}
 of the Galactic center gamma rays gives
strong constraints on the annihilation DM scenario~\cite{Liu:2009sq}.
The decaying  DM~\cite{Ibarra:2008qg,Arvanitaki:2008hq,Yin:2008bs}
with a lifetime at the order $\cal{O}$$(10^{26})s$ is another
elegant way to explain the  cosmic ray anomalies. In particular,
the contraints from the Galactic center gamma rays are much
weaker~\cite{Liu:2009sq}.
However, the ultimate long lifetime of decaying DM
becomes a non-trivial problem since the symmetry, which makes the DM
stable, must be broken tinily.

Supersymmetry naturally solve the gauge hiearchy problem in
the Standard Model (SM). Gauge coupling unification in the
Minimal Supersymmetric Standard Model (MSSM)
implies the Grand Unified Theories (GUTs) at the GUT scale $M_{GUT}$
around $2\times 10^{16}$ GeV. Thus, the DM decays via the dimension-six
operators suppressed by the GUT scale
is a rather simple solution to the long lifetime of decaying
DM, and it may provide a way to probe the GUT scale
physics~\cite{Arvanitaki:2008hq}.
Another problem in the decaying DM is how to understand the DM
dominant leptonic decays,
 especially to the $\mu$ and $\tau$ final states.
Without automatically kinematical suppressions like the
 annihilation models~\cite{ArkaniHamed:2008qn}, one has to employ
 some special symmetry so that the DM interacts strongly with
  the second or third families of the charged leptons,
for instance,  flavor Froggat-Nielson symmetry~\cite{Gao:2010pg}.

Because of the DM leptonic decays,  one may conjecture that
the DM sector only interacts with the lepton sector~\cite{Bi:2009md}.
Note that the neutrino masses are very tiny, we
can parametrize  the small
couplings for the DM decay as the ratio between the light neutrino mass
$m_\nu\sim10^{-11}$ GeV and the GUT scale
\begin{align}
\lambda\sim {m_\nu\over M_{GUT}}\sim 10^{-27}~.~
\end{align}
Interestingly, this is the typical order of tiny coupling parameter rendering
the lifetime around $10^{26}s$ for a TeV-scale DM.
Thus, it also implies the deep connection between
 the decaying DM with long lifetime and the active neutrinos with
tiny masses.

In this paper, we consider the supersymmetric Standard Models with
$R$-parity ($R_p$) and three
right-handed neutrinos (RHNs) $N_i$ where the neutrino masses are generated
via the low-scale seesaw mechanism. In the DM sector, we introduce
the SM singlet DM fields $\phi_{1,2}$ and a discrete $Z_3$ symmetry under which
the term $\phi_1 \phi_2$ is invariant.
 At the leading order they  couple to leptonic sector
through a dimension-seven operator
\begin{align}\label{nnn}
 {1\over \Lambda}\times{1\over M_{N_l}}\times \phi_{1}\phi_{2}\left({{\cal C}'_{ij}\over
M_{N_l}}(L_iH_u)(L_jH_u)\right),
\end{align}
where coefficient ${\cal C}'_{ij}$  come from neutrino Dirac Yukawa couplings,
constrained by light neutrino masses in seesaw mechanism. It can be
obtained after integrating out the heavy right-handed neutrinos
(RHNs) and a superheavy field $X$ with mass $\Lambda$,
 provided that $\phi_i$ only interacts with the RHNs
mediated by $X$. Note that there is a GUT scale in GUTs, we shall assume
$\Lambda\sim M_{GUT}$. Interestingly, this is the exact scale that we need
to produce the correct DM density via the freeze-in mechanism.
In our models, the vacuum expectation values (VEVs) for $\phi_{1,2}$
are close to the RHN masses. Thus,   Eq.~(\ref{nnn}) gives the
small parameter $\lambda$ approximately if we identify the terms in the
bracket as $m_\nu$. Furthermore, the preferred $\mu$ and $\tau$ decay
channels can be related to the  neutrino tri-bimaximal mixing  (TBM) scenario
with the second right-handed neutrino $N_2$ dominant
seesaw mechanism~\cite{King:2005bj}. In short,
cosmic ray anomalies, if confirmed further,
potentially have deep correlation with neutrino physics, especially
such seesaw  mechanism.

The DM relic density generally conflicts with standard
thermal freeze-out scenario in the decaying DM models if the
scalar components of the DM fields acquire VEVs.
It is not difficult to understand it from the effective
operators $\phi_1\phi_2 \mathfrak{L}^2$ ($\mathfrak{L}$ are the operators
for particles in the MSSM) which can generate large annihilating rate.
However,  they also catastrophically make $\phi_i$ decay very fast.
 So one usually  considered the non-thermal DM production,
for example, a detail study given in Ref.~\cite{Bi:2009am}.
Unlike the weakly interacting massive particle (WIMP) scenario,
the non-thermal production  usually  loses  the natural  predication
on   DM  abundance. Recently, it was   proposed
that the  feebly interacting massive particle (FIMP)  may be an
alternative to WIMP~\cite{Hall:2009bx}, shedding  light on decaying
DM. Typically, the FIMP involving a coupling at the strength
$\mathcal {O}(10^{-13})\sim {\rm TeV}/M_{GUT}$ for the decay
dominated freeze-in mechanism, or a larger
 one $\sim\mathcal
{O}(10^{-11})$ for scattering dominated  freeze-in mechanism. Amazingly, in our model,
 $\phi_{1,2}$ must couple to the RHNs  by the
 dimension-five operators suppressed by $M_{X}$ somewhat smaller than
 $M_{GUT}$ and by $M_{N_i}\sim 10^4$ GeV,  based on proper  decaying lifetime of
 DM. Similar results hold for the scattering process  dominated freeze-in
mechanism. Therefore, in our decaying DM model, its  relic density again is
 a ``miracle'' via the freeze-in mechanism.

As a by product in supersymmetric SMs with freeze-in
mechanism, we are able to solve the small scale problem for power spectrum,
in the presence of a metastable $R_p-$odd state $\wt \phi$
in the DM sector. The
point is the following:  the whole supermultiplets $\phi_{1,2}$ are
 freezed into the thermal  bath. We assume that
$m_{\phi_R}+m_{\wt \phi\,'}>m_{\wt \phi}$,
where $\phi_R$ is the lightest state and the DM particle
 while $\wt \phi\,'$
 is the lighter  $R_p-$odd state
and has a mass close to the DM particle $\phi_R$.
By virtue of $Z_3\times R_p$, the
leading decay mode of $\wt \phi$ is to the lightest supersymmetric
particle (LSP) plus $\phi_R$. Provided that $m_{\wt \phi}$  and
the LSP are respectively sufficient
heavy  and light,  the relativistic LSP is produced from $\wt \phi$ late
decay at some sufficient late time $\tau_I\sim 10s-1000s$. This
warm DM component is just the key to reduce power spectrum on small
scale~\cite{Lin:2000qq}.

In addition, we can still explain the baryon asymmetry via soft
leptogenesis~\cite{Fukugita:1986hr} since the seesaw
scale is low around $10^4$ GeV. In
 supersymmetric seesaw framework, when the new  CP-violating
sources in the soft terms dominate the
sneutrino(s) $\wt N$ CP-violating decay, the so-called
  soft  leptogensis~\cite{D'Ambrosio:2003wy,Grossman:2003jv}
is indeed able to produce
enough lepton numbers in our model. Before the discovery of gaugino
effect  \cite{Grossman:2004dz}, soft leptogensis suffers the highly
suppressed bilinear soft mass term $B_{N_i}M_{N_i}\wt {N_i} \wt {N_i}$ with
$B_{N_i}\lesssim10^{-3} M_{SUSY}$ where $M_{SUSY}$ is the
universal supersymmetry breaking scale. In this paper, we assume that the
trilinear soft terms $A Y^{Nij}\wt {N_i} \wt L_j H_u$ are the only sources of
CP-violation in the supersymmetry breaking
 soft terms. Interestingly,  enough
baryon number density  can be generated naturally. Even the
baryon number density tends to be
overproduced if $M_{N_2}$ is too light  $\sim {\cal O}$ (TeV),
we can choose a relatively smaller universal $A$ term  or tune its
CP violation phase to obtain the observed baryon asymmetry.

This paper is organized as follows. In Section II, we present the model,
and discuss the decay of DM, its relic density, as well as  the
relation between neutrino physics and cosmic ray anomalies. In
Section III, we study the phenomenological consequences of our model,
such as the solution to the small scale structure problem,
and the low-scale soft leptogensis. In the Appendix A, we briefly review
the freeze-in mechanism.

\section{The Supersymmetric Decaying Dark Matter Model
with $N_2$ Dominant Seesaw Mechanism}

\subsection {The Decaying Dark Matter Model}

We consider the supersymmetric SM with three  RHNs
and $R$-parity,
and introduce a DM  sector.
As we know, a well defined DM sector
 should not only have a DM particle at the  TeV scale,
but also spontaneously breaks the discrete symmetry that
stabilize the DM particle. The simplest dark sector contains a SM
 singlet $\phi$  and a  $Z_2$ discrete symmetry
under which only $\phi$ is odd. To have a decaying DM,
we break the $Z_2$ symmetry by giving a VEV to $\phi$, ${\it i.e.}$,
 $\lan \phi\ran\neq0$. Thus, at the leading order,
$\phi$ couples to  the observable sector through
dimension-five operators $\phi^2N_i^2/\Lambda$,
which  can be obtained from a renormalizable theory after integrating out
a heavy field with mass $ \Lambda$. This decaying DM can explain
the cosmic ray anomalies and satisfy the other phenomenological requirements
 that we shall consider. For an example, see Ref.~\cite{Matsumoto:2010hz}.
  However, in such an simple model  it
is very difficult to break $Z_2$ symmetry naturally. Concretely
  speaking, we may have to introduce another SM singlet field
 $S'$ that couples to $\phi$ via a superpotential term  $S'\phi^2$.
This superpotential term provides the quartic term to the scalar
 potential of  $\phi$, and then we can realize the  $Z_2$ symmetry
breaking. After $Z_2$ symmetry breaking, $\phi$ and $S'$ will
mix with each other. Because we cannot forbid the direct couplings between
   the SM singlet $S'$ and the observable sector,
this simplest model is excluded unless we have huge fine-tuning.

Therefore, we consider  a DM sector with two SM singlet
fields $\phi_{1,2}$ and a discrete  $Z_3$ symmetry.
Under $Z_3$ symmetry, $\phi_{1,2}$  transform as follows
\begin{align}
\phi_1\rightarrow w \phi_1~,~~~ \quad \phi_2\rightarrow w^2 \phi_2~,~\,
\end{align}
where  $w\equiv e^{i2\pi/3}$. All the  other fields in our model
 are neutral under the $Z_3$ symmetry, so, any  renormalizable coupling
term between the DM sector and observable sector is forbidden
due to the $Z_3$ symmetry and $R$-parity. In particular,  the traditional
particle physics  in the observable sector will not be affected.
The most general $Z_3-$invariant and renormalizable superpotential in
 the DM sector,
as well as  the corresponding supersymmetry breaking soft terms are
\begin{align}\label{}
W_{DM}=&{\lambda_1\over 3}\phi_1^3+{\lambda_2\over
3}\phi_2^3+M_\phi \phi_1\phi_2~,~ \nonumber \\
-{\cal
L}_{soft}^{DM}=&m_{\phi_1}^2|\phi_1|^2+m_{\phi_2}^2|\phi_2|^2+\left({A_{\lambda_1}\over
3}\lambda_1\phi_1^3+{A_{\lambda_2}\over
3}\lambda_2\phi_2^3+B_{\phi}M_\phi \phi_1\phi_2+h.c.\right)~.~\,
\label{darksector}
\end{align}
In fact, this model not only preserves $Z_3-$symmetry, but also a
trivial $R$-parity. In the following,
 we shall prove that this simple DM sector
can break the $Z_3-$symmetry spontaneously, and has a proper
spectrum with a TeV-scale decaying DM coupling to the
observable sector.

As poingted out in the Introduction, to have the desirable
DM lifetime, abundance and decay products,
the DM should couple to the RHNs via the dimension-five operators
suppressed by the GUT scale $\sim M_{GUT}$. This can be achieved by integrating
out a heavy SM singlet field $X$, which mediates the interactions between
the DM sector and observable sector. So we consider the following
superpotential
\begin{align}\label{sup}
W\supset & {M_{N_i}\over 2}N_i^2+Y_{ij}^N L_iH_uN_j
 +{\lambda_{Xi}\over 2} X N_iN_i
\cr& +\lambda_{X\phi
}\phi_1\phi_2X+\left({M_X\over 2} X^2+{\rm irrevelant
\,\,terms}\right),
\end{align}
with $M_X\sim M_{GUT}$. For simplicity, we have  assumed that the RHNs are in
the mass basis.  We can explain the neutrino masses and mixings
by employing some non-Abelian flavor
symmetry such as $A_4$  \cite{Altarelli:2005yp}, although we
do not consider it here. In addition, we do not consider the superpotential
$X H_u H_d$ so that we can explain the PAMELA experiment. This can be
realized in the five-dimensional scenario compactified on
$S^1/Z_2$ (or in the M-theory on $S^1/Z_2$) where $X$ and $H_u/H_d$ are localized
on the different D3-branes on the two boundaries of $S^1/Z_2$ while the
right-handed neutrinos are in the bulk.

To obtain the effective action below the scale $M_X$, we
 integrate out the heavy field $X$ through its equation of motion
\begin{align}\label{}
M_XX+\lambda_{X\phi}\phi_1\phi_2+\lambda_{Xi}N_i^2 ~=~0~.~\,
\end{align}
So we obtain the desirable dimension-five operators, which
 describe the interactions between the DM sector and
 RHNs. The effective superpotential are
\begin{align}\label{WN}
W_{N,eff}=&W_{hid} -{\lambda_{X\phi}\lambda_{Xi}\over
2M_X}\phi_1\phi_2N_i^2\cr &+\left({M_{N_i}\over
2}N_i^2+Y_{ij}^NL_iH_uN_j\right)+(...)~,~
\end{align}
where dots denote the irrelevant corrections after
 integrating out $X$. Also,
 the corresponding supersymmetry breaking
soft terms are  given by
\begin{align}\label{softterm}
-\mathcal {L}_{soft}\supset&\mathcal {L}_{soft}^{hid}+m_{\wt
N_i}^2|\wt N_i|^2\cr &+  \left({{\cal C}^\phi\over 2} A_{\phi
N_i}\phi_1\phi_2N_i^2+A_{ij}Y_{ij}^N \wt L_iH_u{\wt
N}_j+{B_{N_i}\over 2}M_{N_i}{\wt N}_i^2+h.c.\right),
\end{align}
where
\begin{align}
{\cal C}^\phi\equiv - {\lambda_{X\phi}\lambda_{Xi}\over M_X}~.~\,
\end{align}
Especially, the effective action described by Eqs.~(\ref{WN}) and
(\ref{softterm}) will provide the dynamics to freeze-in the DM
particles, and generate baryon asymmetry at the
scale around $ M_{N_i}$.

\begin{figure}[htb]
\begin{center}
\includegraphics[width=3.3in]{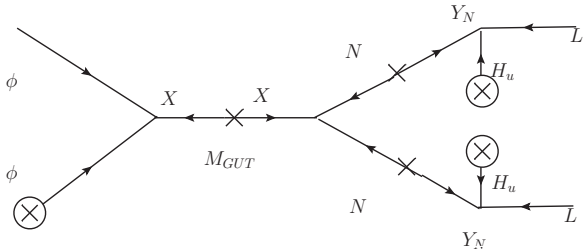}
\end{center}
\caption{\label{d=6}  Feymann diagram for the dimension-7  operators
 ${\cal C}_{ij}\phi_1\phi_2(L_iH_u)(L_jH_u)/M_XM^2_{N_l}$
generated by integrating out $X$ and $N_i$
at tree level.}
\end{figure}

To obtain the  effective action below the right-handed neutrino
mass scale, we further
integrate out the RHNs through their equations of motion
\begin{align}\label{}
M_{N_j}N_j+Y_{ij}^NL_iH_u -{\lambda_{X\phi}\lambda_{Xi}\over
M_X}\phi_1\phi_2N_i~=~0~.~\,
\end{align}
Then,  the leading order operators coupling $\phi_i$ to the SM particles
are dimension-seven operators presented in the Introduction,  ${\cal
C}_{ij}\phi_1\phi_2(L_iH_u)(L_jH_u)$ (also see Fig.~\ref{d=6}),
which can account for cosmic ray anomalies. The
operator  coefficients are
\begin{align}\label{}
{\cal C}_{ij}=-{\lambda_{X\phi}\lambda_{N_l}\over
M_X}{Y_{il}^NY_{jl}^N\over M_{N_l}^2}~,
\end{align}
where  $l$ is  summed over three RHNs. Those
coefficients have a clear correlation with neutrino Dirac Yukawa
couplings as well as the light neutrino Majorama mass matrix.
After $Z_3$ symmetry breaking by the VEVs of $\phi_i$, we
obtain the renormalizable
interactions between the DM and (s)leptons from superpotential terms
$\phi \nu_i \nu_j $, as well as the dimension-five interactions $\phi \nu_i (L_jH_u)$.
The dimension-five operators are interesting since the DM
particles can decay dominantly to the $\mu$ and $\tau$ leptons due to
the neutrino TBM.

\subsection{Spontaneously  $Z_3$ Symmetry Breaking and Decaying DM}

Cosmic ray anomalies can be explained elegantly by the long-lived decaying
DM with lifetime $\tau\sim 10^{26}$s that decay dominantly to the
charged leptons. Because the DM lifetime is so long, it is natural to
have a symmetry if the DM is stable.
In our model, this symmetry is the discrete $Z_3$ symmetry.
To break the $Z_3$ symmetry spontaneously, we consider the
 relevant scalar potnetial $V(\phi_i)$ from Eq.~(\ref{darksector})
\begin{align}\label{poten}
V(\phi_i)=&|\lambda_1
\phi_1^2+{M_{\phi}}\phi_2|^2+|\lambda_2\phi_2^2+M_\phi\phi_1|^2\cr
&+m_{\phi_1}^2|\phi_1|^2+m_{\phi_2}^2|\phi_2|^2+\left({A_{\lambda_1}\over
3}\lambda_1\phi_1^3+{A_{\lambda_2}\over
3}\lambda_2\phi_2^3+B_{\phi}M_\phi \phi_1\phi_2+h.c.\right) .
\end{align}
Note that $M_X\gg  M_{\phi_i}$, the contributions to
the low energy effective scalar potential from
the superpotential $X \phi_1 \phi_2$ are very small,
and then we do not consider them. Because
 Eq.~(\ref{poten}) contains quite a few parameters, analytical
study is pretty difficult.  To reduce the parameters in the DM sector,
we assume that the squared
soft masses $m_{\phi_i}^2$ are universal, and
the trilinear soft terms  $A_{\lambda_i}$ are universal. Moreover,
to avoid the Landau pole problem for Yukawa couplings below the GUT
scale, we choose $\lambda_1=\lambda_{2}=0.3$.

First, we parametrize the fields $\phi_i$ as follows
\begin{align}
\phi_{i}=v_i+{{\phi^0_{i,R}}+i\,{\phi^0_{i,I}}\over \sqrt{2}},
\label{VEVs-phi}
\end{align}
where ``0'' denotes the interaction eigenstates.
We require that the spectrum have the following
 properties: (i) The lightest scalar as the DM particle should be
  about 2 TeV from the Fermi-LAT electron excess at high energy
region. (ii) There should be a heavy and sufficient long
 lived $R_p-$odd fermion  with mass about 5 TeV so that
we can solve the small scale problem.
Although these requirements impose some
constraints on parameter space, they can still be satisfied easily.
In the following, we present an explicit  example
whose input parameters are
\begin{align}\label{v}
&\lambda_1=\lambda_{2}=0.3,\quad M_\phi=1.0 {\rm\, TeV},\cr &
 m_{\phi_i}^2=200{\,\rm GeV^2},\quad B_\phi=
-600{\rm\, GeV},\quad A_{\lambda_i}=600 {\,\rm GeV}.
\end{align}
Let us comment on the above choice of parameters.
With the fixed $\lambda_i$, without tuning on
supersymmetry breaking soft terms,  larger  $M_\phi$ will
generate larger VEVs for $\phi_i$ as well as heavier spectrum,
which is disfavored by the DM mass requirement. Because $A_{\lambda_i}$
have the same sign, we choose negative $B_\phi$ so that
the VEVs for $\phi_i$ have the same sign as well and
we can have an absolute stable vacuum.
The larger $A-$terms  help us to have a more
phenomenologically interesting spectrum, $i.e.$, to increase  the
mass splitting between $\wt\phi$ and $\phi_R$ and meanwhile to keep
$m_{\wt \phi}<m_{\phi_R}+m_{\wt\phi\,'}$.

Numerically, one  global minimum is located at
\begin{align}
& \langle \phi_1 \rangle \equiv v_1\approx -6.06{\rm\, TeV},\quad
\langle \phi_2 \rangle \equiv  v_2\approx -6.57{\rm\, TeV}.
\end{align}
At this vacuum,  the various  mass eigenvalues and corresponding
eigenstates  are respectively given by
\begin{align}
&m_{\phi_R}\approx 2.60 {\rm\, TeV},\quad m_{\phi_R'}\approx
4.81{\rm\, TeV}, \cr
 &m_{\phi_I}\approx 2.91 {\rm\, TeV},\quad
m_{\phi_I'}\approx 4.83{\rm\, TeV},\cr & m_{\wt \phi}\approx
-4.80{\rm\, TeV}, \quad  m_{\wt
\phi\,'}\approx -2.78 {\rm\, TeV}.\\
&\phi_R=0.70\phi^0_{1,R}+0.71\phi^0_{2,R},\quad
\phi_R'=0.71\phi^0_{1,R}-0.70\phi^0_{2,R},\cr
&\phi_I=0.82\phi^0_{1,I}+0.58\phi^0_{2,I},\quad
\phi_I'=-0.58\phi^0_{1,I}+0.82\phi^0_{2,I},\cr &\wt\phi=-0.65\wt
\phi_1^0+0.76\wt\phi_2^0,\quad \wt\phi\,'=0.76\wt
\phi_1^0+0.65\wt\phi_2^0\label{eigenphi}.
\end{align}
Thus, the lightest CP-even state $\phi_R$ is the DM particle accounting
for the cosmic ray anomalies. $\phi_I$, $\wt \phi\,'$
and $\phi_{I,R}'$ are unstable but will contribute to the DM abundance
 in the freeze-in mechanism. The heavy metastable state $\wt\phi$
is crucial to solve the small scale problem on power spectrum.
Notice that we have arranged parameters to have $m_{\wt
\phi}<m_{{\wt \phi}\,'}+m_{\phi_R}$,  $\wt \phi$ can not decay to
${\wt \phi}\,'$ and $\phi_R$ at two-body level. We emphasize that
with suitable mass  $\wt \phi\,'$  might also constitute   a
component of DM today by forbidding its two body-decay to $\phi_R$
and  gravitino $\wt G$. By the way, all the mixing factors are
nearly democratic about $ 0.7$, so for simplicity we may drop this
factor in the following discussions, and the subscripts for $\phi_i$
may be ignored since they will neither  affect the discussions nor
bring any misunderstanding.

After the $Z_3$ symmetry breaking, the lightest  $Z_3-$odd state
is unstable and
decays to leptons through the heavy RHNs, which  is described in
 Fig.~\ref{d=6}. The DM particles can decay via the operators ${\cal
C}_{ij}\phi_1 \phi_2 (L_iH_u)(L_jH_u)$, and
 we are interested in the final sates
containing $\mu$ and $\tau$. At the leading order, such DM decays
are described by the dimension-five operators ${\cal C}_{ij}^5\phi
\nu_i(L_jH_u)$ obtained from dimension-seven
operators with  one VEV for $\phi_i$  and
one VEV for  $ H_u$. To show the close relation between DM decay
and neutrino masses/TBM, we express  the dimension-five operator coefficients
into the  light neutrino mass matrix elements.  First,
the Dirac neutrino mass matrix can be written as
\begin{align}\label{LR}
M_{LR}=Y^N\langle H_u^0\rangle=v\sin\beta\times({\cal N}_1,{\cal
N}_2,{\cal N}_3),
\end{align}
where ${\cal N}_i$ is the $i$-th column, and $\tan\beta=\langle
H_u^0\rangle/\langle H_d^0\rangle$ as in the MSSM,
$v= {\sqrt {(\langle H_u^0\rangle)^2+ (\langle H_d^0\rangle)^2}}=174$  GeV.
Using the seesaw formular  $M_{LL}= M_{LR} M_{RR}^{-1}M_{LR}^T$,
 we get  the coefficients
\begin{align}\label{c5}
{\cal C}_{ij}^5=& - {\lambda_{X\phi}\over
M_X}\left({v^2\sin^2\beta{\cal N}_l{\cal N}_l^T\over
M_{N_l}}\right)_{ij}{{\lambda_{Xl}}\,v_\phi\over v\sin\beta M_{N_l}}
\cr
 \approx
&
 - {\lambda_{X\phi}\over v\sin\beta}
 \left({(M_{LL})_{ij}\over M_X}\right)
 \left({v_\phi\over
 M_{N_2}}\right),
\end{align}
where  $v_\phi$ denotes  $v_{1}$ or $v_2$, and the family universal couplings
$\lambda_{Xl}\simeq1$ are assumed. This approximation is valid if
$M_{N_2}$ dominates the seesaw contributions to $M_{LL}$ and
$M_{N_2}\sim M_{N_1}$. Thus, the DM decays are closely related to the
light neutrino mass matrix (elements), which will be studied
in the next Section. We will show that the entries in the light
neutrino mass matrix  $(M_{LL})_{ij}$ are at the same order
(about the heaviest neutrino mass) for $i,~j=2,~3$, while the
other entries are much smaller (around the second
 heaviest neutrino mass or smaller).

The DM branch decay lifetime is
\begin{align}\label{life}
\tau({\phi}_R\rightarrow \wt \nu_i\ell_j\wt H_u)&\approx
768\pi^3\times {1\over ({\cal C}_{ij}^5)^2} {1\over m_{\phi_R}^3}
\cr & =3.6\times10^{26}\times \left({M_X\over 10^{15}\,\rm
 GeV}\right)^2\times\left({0.05\,{\rm eV}\over
(M_{LL})_{ij}}\right)^2\times\left({M_{N_2}\over 10^4\,\rm
 GeV}\right)^2 \cr & \times\left(5\,{\rm TeV}\over
v_{\phi}\right)^2\times\left(2\,{\rm TeV}\over m_{\phi_R}\right)^3s,
\end{align}
where we have taken $\tan\beta=5$ throughout this paper. The
actual  lifetime  does not depend on it much since a larger
$\tan\beta$ always gives $\sin\beta\approx 1$. We keep
$\lambda_{X\phi}$ as an adjustable parameter to obtain
the proper lifetime of $\phi_R$,  which will be chosen
as $0.5$ from then on. In order to generate the DM density
via freeze-in mechanism, we choose $M_{N_i}/M_X\sim 10^{-11}$.
And then we explain the neutrino masses and mixings via
the low-scale seesaw mechanism. Therefore,
as pointed out in the
Introduction, the crucial point to get such a long lifetime decaying
DM  is the combined factor $ M_{LL}/M_X \sim 10^{-26}$.

\subsection{$N_2$ Dominant Seesaw Mechanism and Cosmic Ray Anomalies}

Although our model can generate the suitable DM lifetime naturally,
the dominant  decays to the leptonic final states and the fittings of
the PAMELA and Fermi-LAT data  need  further study.
Especially, the decaying product should be dominated by the second
and third  families of charged leptons~\cite{Meade:2009iu}.
Note that the approximate $\mu-\tau$ symmetry is
introduced to explain the light neutrino masses and
mixings~\cite{Fukuyama:1997ky}, we suggest that the DM
decay is related to the $N_2$ dominant seesaw mechanism which
can explain neutrino TBM~\cite{King:2005bj}.

With  approximate $\mu-\tau$ symmetry~\cite{Fukuyama:1997ky},
we obtain  the general light Majorana mass
matrix  by four parameters
\begin{align}\label{mutau}
M_{LL}=m_0\left(\begin{array}{ccc}
         X&Y&Y\\
Y&Z&W\\
Y&W&Z
      \end{array}\right)~.~
\end{align}
It predicts the maximal atmosphere mixing angle $\theta_{23}=\pi/4$
and $\theta_{13}=0$, but leave the solar mixing angle
$\theta_{12}$ arbitrary. Taking $\sin^22\theta_{12}=8/9$,
the neutrino TBM is
obtained~\cite{Harrison:2002er}. The TBM $M_{LL}$ mass matrix
only has three parameters since this fixed $\theta_{12}$ is
equivalent to a relation $Z+W=X+Y$. So we have
\begin{align}\label{TBM}
M_{LL}=m_0\left(\begin{array}{ccc}
         X&Y&Y\\
Y&X+V&Y-V\\
Y&Y-V&X+V
      \end{array}\right).
\end{align}
In the framework of seesaw mechanism with heavy RHN
dominance, the crucial point of neutrino mixings  is the specially
aligned Dirac neutrino mass matrix (or say the Yukawa coupling matrix).
Concretely speaking, the neutrino TBM can be understood by the
aligned vacuum from an $A_4$ discrete flavour
symmetry breaking~\cite{deMedeirosVarzielas:2005qg}.

To explain why the DM decays dominant to muon and tau via neutrino
physics, we modify  the original Dirac Yukawa coupling
 ansatz used in  Ref.~\cite{King:2005bj} as follows
\begin{align}\label{MLR}
M_{LR}=\left(\begin{array}{ccc}
         A&\quad 0& \quad 0\\
A&\quad -B& \quad0\\
A&\quad B&\quad C
      \end{array}\right)~.
\end{align}
 To produce the realistic neutrino masses and mixings, we assume three
RHNs  with proper mass hierarchy $M_{N_1}\lesssim M_{N_2}\ll
M_{N_3}$ so that the light neutrino mass matrix
accommodates both  the TBM and the $\mu+\tau$ dominated decay product of
DM. Thus, the light neutrino mass matrix  is
\begin{align}
M_{LL}&=v^2\sin^2\beta\left({{\cal N}_1{\cal N}_1^T\over
M_{N_1}}+{{\cal N}_2{\cal N}_2^T\over M_{N_2}}+{{\cal N}_3{\cal
N}_3^T\over M_{N_3}}\right) \cr &=m_0\left(\begin{array}{ccc}
         X&Y&Y\\
Y&X+V&Y-V\\
Y&Y-V&X+V
      \end{array}\right)+{\cal O}(C^2/M_{N_3})~,
\label{MLL-nu}
\end{align}
where $X=Y=A^2/(M_{N_1}m_0)$, and $V=B^2/(M_{N_2}m_0)$. The
last term gives the subdominant contributions to $M_{LL}$, but
it is still important for the  mass of the
lightest neutrino.

Now we show that the DM dominant decay channel to $\mu+\tau$ is
a natural result for the $N_2$ dominant seesaw mechanism if
the neutrino masses are  normal hierarchy. Combining the
DM decays with neutrino masses and TBM gives some contraints
on the free parameters.
First,  it is obvious that $M_{LL}$ should be in the second RHN dominance.
Next, with Eq.~(\ref{MLL-nu}) we obtain
three neutrino  approximate masses
\begin{align}\label{nmass}
m_{\nu_3}\approx 2Vm_0={2B^2\over M_{N_2}},\quad m_{\nu_2}\approx
3Xm_0={3A^2\over M_{N_1}}, \quad m_{\nu_1}\lesssim {\cal O}\left(
{C^2\over M_{N_3}}\right),
\end{align}
in a normal hierarchy form. Thus, the neutrino oscillation data
$\Delta m_{21}^2\approx 7.65\times 10^{-5}$ eV$^2$ and $\Delta
m_{31}^2\approx 2.40\times 10^{-3}$ eV$^2$ suggest that
\begin{align}\label{nmass}
{B^2\over M_{N_2}}: {A^2\over M_{N_1}}\simeq 8.4:1~,~
\end{align}
is valid when the $N_3$ is sufficient heavy~\cite{Schwetz:2008er}. But from
Eq.~(\ref{c5}), the dimension-five operator coefficients are
proportional to $1/M_{N_l}^2$. And then  they disfavor large hierarchy
$M_{N_2}\gg M_{N_1}$. So the hierarchy in Eq.~(\ref{nmass}) is mainly due
to the moderate relation  $A<B$. Because $B^2/M_{N_2}$ will appear
several times later,  we fix it in the massless $\nu_1$ limit
(or say infinite $m_{N_3} $ limit)
\begin{align}\label{B}
{B^2\over M_{N_2}}\approx \sqrt{\Delta m_{31}^2/2}\approx
0.035\,{\rm eV}.
\end{align}

The DM $\phi_R$ decays to the SM fermions  are the dominant  primary source
of comic ray since the lifetime of  other states such as  $\wt \phi$
is much short at the cosmic time  scale. At tree level,  the $\phi_R$
three-body decay modes are
\begin{align}\label{decay}
  \phi_R\rightarrow \ell_i H_u \nu_j,\quad \wt \ell_i \wt H_u \nu_j,
\end{align}
and the corresponding lifetime estimation is given in
Eq.~(\ref{life}). In Ref.~\cite{Gao:2010pg}, it has been
explicitly simulated the  electron spectra, and found  that
the spectra from direct hard leptons plus the soft contributions from
cascade decays via the slpeptons and Higgs, Higgsinos are able to fit
the PAMELA and Fermi-LAT experiments
while not produce the anti-proton excesses in the constrained MSSM.
Moreover, for the  two-body decay modes,  $\phi_R$ decays to pure
(two) neutrinos. The branch decay lifetime is approximately given by
\begin{align}\label{2bodylife}
\tau({\phi}_R\rightarrow \nu_i\nu_j)&\approx 8\pi\times {1\over
({\cal C}_{ij}^5)^2} {1\over m_{\phi_R}^3}\left( {m_{\phi_R}\over
v\sin\beta}\right)^2,
\end{align}
which  is about 20$\%$ of the one through three-body decays. Assuming
a lifetime about $5\times 10^{26}$s  for three-body decay modes to
explain the comic ray anomalies, we have  $\tau({\phi}_R\rightarrow
\nu_i\nu_j)\sim 10^{26}$s. The produced  neutrino signals are
potentially detectable with the upcoming IceCube neutrino observatory~\cite{IceCube},
and the constraints on the DM models can be found
in Ref.~\cite{Buckley:2009kw}.

\subsection{DM Density from Freeze-in Mechanism}

The decaying DM abundance can be produced naturally
via the freeze-in mechanism (for a brief review, see
Appendix~\ref{freezein}) in our models.
Let us explain the cosmological  setup first since
it is important
to make freeze-in mechanism work. The initial relic density of
$\phi_i$ should  almost vanish, while (s)RHNs have the thermal
density  at least at $T\sim M_{N}$. However, for the low-scale seesaw
mechanism in
the MSSM, the (s)RHNs only weakly interact with the plasma
because the neutrino Dirac Yukawa couplings are small about
$10^{-5}$. Thus,  after inflation the MSSM particles
in the plasma alone cannot   produce the thermal (s)RHNs.
But this problem  can be
solved easily in the Next to the
MSSM (NMSSM), where the superpotential term $S N_i^2$
can be introduced. Also, the suitable density
for the (s)RHNs can be  produced non-thermally
by  coupling them directly to the inflaton field. Thus,  we assume
the (s)RHNs   in the plasma at the temperature $T\gtrsim
M_{N_i}$. But the initial densities for $\phi_i$ are ignorable
since they are SM singlets and only very weakly
interacts with the (s)RHNs.

 However, during the decoupling of $\wt
N_i$,  in the absence of inverse decay, the tiny branch decays
or the scattering processes of $\wt N_i$ and $N_i$
produces $\phi_i$. To have the natural relic
densities of $\phi_i$ via freeze-in mechanism,
the typical couplings are required to be around
 $ 10^{-13}$ for two-body decays and
$10^{-11}$ for two to two scattering processes~\cite{Hall:2009bx}.
 To be concrete, we give the relevant terms
 between (s)RHNs and $\phi_i$/$\wt \phi_i$
for freeze-in mechanism
\begin{align}\label{freezeinL}
-{\cal L}\supset &{\cal C}^\phi(\phi_1\wt \phi_2+\phi_2\wt \phi_1)
\wt N_i N_i+{{\cal C}^\phi\over 2}\wt \phi_1 \wt \phi_2 \wt
N_i^2+|F_{\phi_1}|^2+|F_{\phi_2}|^2+|F_{N_i}|^2,\cr
 \rightarrow
 &
{\cal C}^\phi(v_1\wt \phi_2+v_2\wt \phi_1) \wt N_i N_i
 +
 {\cal C}^\phi M_{N_i}\left(\phi_1\phi_2\wt N_i\wt
 N_i^*+h.c.\right)+(...),
\end{align}
where we only consider the dominant terms, and dots denote  many
ignored terms, such as the supersymmetry breaking
 trilinear soft terms since the freeze-in
amplitudes are controlled by $M_{N_i}\gg A_{ij}$. Similarly,  the
scattering processes are also sub-dominated
sinces they are proportional to $v_i$ which are
several times smaller than $ M_{N_i}$ in our model. In the precise
calculations, one has to transform the interaction eigenstates to the
mass eigenstates. For the DM state transformations, please see
Eq.~(\ref{eigenphi}). In the following analysis we shall show
that the DM relic density  can indeed be obtained through the
freeze-in mechanism, and the order-one mixing factor is not considered
for simplicity. The mass eigenstates for $\wt N_i$ are
\begin{align}\label{eigenN}
\wt N_{i,\pm}={1\over \sqrt{2}}(\wt N_i\pm \wt N^*_i),
\end{align}
where the squared mass eigenvalues are respectively given by $M_{\wt
N_{i,\pm}}^2=M_{N_i}^2+m_{\wt N_i}^2\pm B_{N_i}$. Here,  $m^2_{\wt N_i}$
are the soft mass square for the  sRHNs.

First,  the freeze-in FIMPs from $\wt N_{i,+}$ and $N_i$
two-body decays  $\wt N_{i,+}\rightarrow  N_i \wt \phi$ and
$N_i\rightarrow \wt N_{i,-} \wt \phi$ as well as $\wt
N_{i,+}\rightarrow \wt N_{i,-}  \phi_{IR}$
 are in general kinetically forbidden.
 In fact, in the natural soft mass scale around $ {\cal O}$(1~TeV),
 the mass splittings among  $\wt N_{i,+}$, $\wt N_{i,-}$ and $N_{i}$
 are about $ M_{SUSY}^3/M_{N_i}^2$, which at most are tens of GeVs.
Consequently there are no decay channels. In fact,
it is  required  for proper DM relic density since the typical couplings
given above are $\sim M_{N_i}/M_X \gg 10^{-13}$. In short,
 these  two-body decays must be forbidden (or at least suppressed sufficiently),
 otherwise, the freeze-in mechanism tends to  over freeze DM(s)
into the  plasma. We have to point out that the above conclusion holds
only when there are  no mixings among the RHNs. If there
exist the mixings $\epsilon_{ij}$ in a complete
 model, we have to require that $\epsilon_{ij}{{\rm Max}\{M_{N_i},M_{N_j}}\}/M_X\sim
 10^{-13}$ since the mass splittings between the RHNs should be  small enough
 to suppress the transition between $N_i$ and $N_j$ significantly.

The scattering process $\wt N_i \wt N_i\rightarrow \phi_{IR}\phi_{IR}$
from the second line of Eq.~(\ref{freezeinL})  can produce
the phenomenologically important components $\phi_{IR}$.
The scattering process $\wt N_i N_i\rightarrow \wt \phi\phi_{IR}$ can be
studied similarly, so we will not present it in this paper.
Numerically, the exact coincidence
is a result of the dimension-five operators $\phi_1 \phi_2 N_iN_j/M_X$.
Note that the RHNs have masses about 10 TeV, and $M_X$ can be
chosen a little bit smaller than $M_{GUT}$, we obtain
 $ M_{N_i}/M_X\sim 10^{-11}$ at the desired order. For the
scattering processes,  we consider the interaction
eigenstates as the mass eigenstates for simplicity
since the mixings are
democratic. The total cross sections are simply given by
\begin{align}
\sigma(\wt N_i \wt N_i\rightarrow \phi\phi)={\lambda_{N_i}^2\over
8\pi s}\lambda^{-1/2}(1,x_{\wt N_i},x_{\wt N_i})~.~\,
\end{align}
The function from phase space $\lambda(a,b,c)\equiv (a-b-c)^2-4bc$
($x_I\equiv m_I^2/s$)  implies that only the lighter species (at
least lighter than $\wt N_i$) could  be freezed into plasma with
significant number densities.  Numerically, the integral factor
$\mathcal {I}[x,z]$ in Eq.~(\ref{22S}) is about 0.5, so the relic
densities of $\phi_{IR}$ and $\wt \phi$ are estimated to be
at the right order
\begin{align}
\Omega_{\phi} h^2\sim& {6.0\times 10^{22}g_{\wt {N}_i}^2 \over g_*^S
\sqrt{g_*^\rho}} \left({m_{\phi}\over
M_{N_i}}\right)\lambda_{{N_i}}^2\cr =&0.065\left({m_{\phi}\over
2.5{\rm TeV}}\right)\left({10 {\rm TeV}\over
M_{N_i}}\right)\left({229^{3/2}\over
g_{*MSSM}}\right)\left(\lambda_{N_i}\over{5\times
10^{-11}}\right)^2,
\end{align}
where $m_\phi$ denotes the mass of $\phi_{IR}$ or
$\wt \phi$. The relic density of gravitino
$\wt G$, which comes from the non-thermal
production via the $\wt \phi$ late decay, is about one order smaller
than $\Omega_{\phi_{R}}$ due to the
mass ratio $m_{\phi_{R}}/m_{\wt G}\sim 10$. We emphasize that
this is not the final DM relic density, and the actual DM density is
obtained by calculating all the processes with exact mixing factors.
However, our results are enough to show that we can generate the
correct DM relic density in our parameter space.

\section{Phenomenological Consequences}

\subsection{Warmed $\wt G$  and Small Scale Problem}

In our model, both $\phi_R$ and $\wt \phi$ are generated
with equal number densities via the freeze-in mechanism.
Because $\wt \phi$ is a relatively heavy  metastable
$R_p-$odd state, it will decay and produce some relativistic
particles. Thus, we can solve the small scale problem
on power spectrum if the relativistic particle is the
 DM candidate like the LSP in the MSSM.
If the comoving free-streaming scales of the relativistic
particles, {\it i.e.}, their motion in the comoving framework from
their production
time $t_I$ till to the matter and radiation equality era $t_{EQ}\approx
2.2\times10^{12}s$, can reach
 the small scale $\cal{O}$(0.1) Mpc,  the power spectrum on
 small scale can be reduced~\cite{Colin:2000dn}. Such warm DM scenario
was proposed in Ref.~\cite{Lin:2000qq}.

To solve the small scale problem in our model, we require
that $\wt \phi$ have a  proper mass (about $ 5$ TeV in
our example parameters) and a proper lifetime.
Lifetime is fine in our model.  Since $\wt \phi$ is odd under
$Z_3\times R_p$, its leading decay mode is given by (as mentioned
previously, in our  interesting  parameter space $\wt \phi$ can not
decay to ${\wt \phi}\,'$ and $\phi_R$ at two-body level)
\begin{align}
\wt \phi\rightarrow \wt G+\phi_R\rightarrow ...~.~\,
\end{align}
Because the decay rate is suppressed by $1/M_{\rm Pl}^2$ in gravity mediation,
the $\wt \phi$ lifetime can be sufficiently  long about $10-1000$ s,
and can be even longer depending  on mass splitting between the bosonic
and fermionic states. However, if $\wt G$ is not the LSP, the above decay chain
will produce a LSP such as  neutralino.
Thus, we have two viable solutions: (i)
$\wt G$ itself is the LSP with mass  about $ {\cal O}(100)$ GeV,
then it is warmed enough to reduce power spectrum on small scale;
(B) $\wt G$ is not the LSP and  has mass $\gtrsim {\cal O}$(TeV).
We require that $\wt G$ can
produce the warm LSP via its decay while not forbid the two-body
decay of $\wt \phi$. In short, these viable solutions do
 not conflicts with the parameter space in the  MSSM.

Because the  late decay of
$\wt G$  may spoil the successful predication of big bang primary
nucleosynthesis (BBN), we consider
 $\wt G$ as the LSP for simplicity.
 In fact, this process can be regarded as a
method of non-thermal production of $\wt G$.  The comoving
free-streaming scale of a freely propagating particle can be calculated from
the formular~\cite{Lin:2000qq}
\begin{align}\label{freestr}
R_f=&\int_{t_I}^{t_{EQ}}{v(t')\over a(t')}dt'\cr \simeq&
2v_0t_{EQ}(1+z_{EQ})^2\log\left(\sqrt{1+{1\over
v^2_0(1+z_{EQ})^2}}+{1\over v^2_0(1+z_{EQ})}\right),
\end{align}
where $z_{EQ}$ and $t_{EQ}$ are the red shift and comic time at the
matter-radiation equality era. Also, $v_0$ is the current velocity of $\wt G$
\begin{align}\label{v0}
v_0={T_0\over T_I}{E_I\over m_{\wt G}},
\end{align}
where $T_0\approx 2.73$ K, and $E_I$ and $T_I$ are respectively the
energy and  temperature when warm $\wt G$ is produced. According
to Eq.~(\ref{freestr}), in order to explain the small-scale
structure,  $v_0$  should take the value
$10^{-8}-10^{-7}$~\cite{Lin:2000qq}.

If $\wt G$ is light, $v_0$ is not dependent on its mass $m_{\wt G}$.
Thus, with the proper mass for
$\wt \phi$ in our example,  we can indeed solve the small scale problem.
Let us  explain it in details. The two-body decay rate of $\wt\phi$ to
its partner $ \phi_R$
 plus gravitino is calculated to be~\cite{Martin:1997ns}
\begin{align}\label{gravi}
\Gamma_I={1\over 48\pi}{m_{\wt \phi}^5\over m_{\wt
G}^2M_{\rm P}^2}\left\{1-\left(m_{ \phi_R}\over m_{\wt
\phi}\right)^2\right\}^4,
\end{align}
with  the reduced Planck mass  $M_{\rm P}\equiv M_{\rm Pl}/\sqrt{8\pi}\simeq
2.4\times 10^{18}$ GeV. Notice that this result is only valid when
gravitino is much lighter than the mass splitting between particle
and its superpartner, this is the exact situation needed in our model:
 $\wt G$ has a large velocity (warm enough) when it was produced.
According to Eq.~(\ref{gravi}),  the cosmological temperature (in
the radiative dominant era) is given by
$T_I=\left(0.301g_*^{-1/2}M_{\rm Pl}/t_I\right)^{1/2}$, where $g_*$ is
the effective relativistic degree of freedom in the plasma, and
$t_I=1/\Gamma_I\sim {\cal O} (10) s$ with $m_{\wt G}\sim 100$ GeV.
Then  we can parametrize  $v_0$ as follows
\begin{align}\label{v00}
v_0=1.8\times 10^{-8}\times  \left({g_*\over 3.36}\right)^{1/4}
\left( {m_{\wt \phi}\over 5{\,\rm TeV}}\right)^{3/2}  \left( {\Delta
m^2_{\phi}\over 20{\,\rm TeV^2}}\right),
\end{align}
where $\Delta m_\phi^2\equiv m_{\wt \phi}^2- m_{\phi_R}^2$, and we have
used $E_I=m_{\wt \phi}/2\left(1-m_{ \phi_R}^2/m_{\wt
\phi}^2\right)\approx m_{\wt \phi}/2$. As pointed out at the beginning,
$v_0$ does not  depend on $m_{\wt G}$ explicitly, and the solution
to the small-scale problem only depends on the mass of $\wt \phi$.

\subsection{Baryon Asymmetry via  Soft Leptogenesis}

Interestingly, we can also explain  baryon
asymmetry via  the  soft leptogensis, since the proper decay
rate of $\phi_R$ requires a low-scale seesaw mechanism with
$M_{N_i}\sim 10^{4}$ GeV at least for  $i=1,~2$.
Consequently, the new lepton number violation and CP violation in the
supersymmetry breaking soft terms play major role in  soft leptogensis.
The soft leptogensis can work only well below
$M_{N_i}<10^9$~\cite{Grossman:2003jv,D'Ambrosio:2003wy}, which is
the low bound on the right-handed neutrino mass
in thermal leptogensis. And then reheating temperature is well below
$10^9$ GeV as well. Thus, we can
reduce the gravitino density produced by thermal scatterings in the
thermal  bath,  and then the $\wt G$ late decay will not destroy
BBN~\cite{Giudice:2008gu}. In short, the possible gravitino problem
in the thermal leptogenesis can be solved.
For a complete review, please see Ref.~\cite{Davidson:2008bu}.
In our model, the relevant
Lagrangian  for soft leptogenesis is
\begin{align}\label{lepto}
-{\cal L}\supset& {1\over \sqrt{2}}\wt
N_{2+}(Y_{i2}^N)^*(M_{N_2}+A_{i2}^*)\wt L_i^\dagger
H_u^\dagger+{1\over \sqrt{2}}\wt
N_{2-}(Y_{i2}^N)^*(M_{N_2}-A_{i2}^*)\wt L_i^\dagger H_u^\dagger\cr
&+(Y_{i2}^N)^*(\wt N_{2+}-\wt N_{2-})L_i^\dagger\wt
H_u^\dagger+{1\over 2}M_2\wt \lambda_2\wt \lambda_2+h.c.,
\end{align}
where $A_{i2}=|A_{i2}|e^{i\theta_{A_{i2}}}$, and  all the other soft
terms have been taken  real. Moreover, the $SU(2)_L$ gaugino mass
$M_2$ is assumed to be real so that it  will not induce large CP violation in
the MSSM.

\begin{figure}[htb]
\begin{center}
\includegraphics[width=4in]{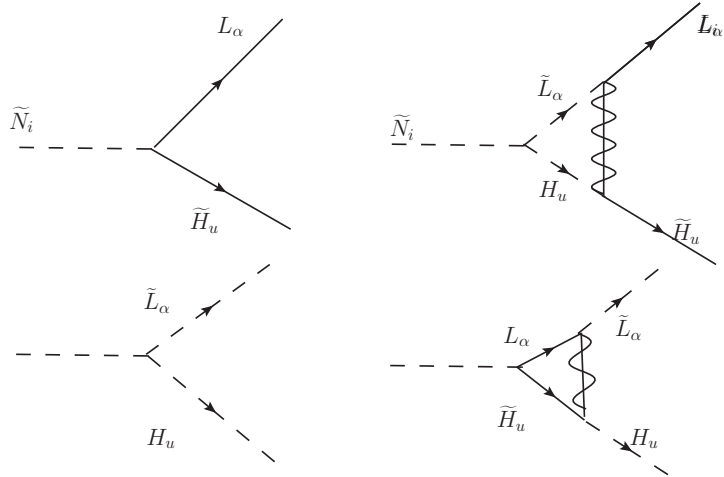}
\end{center}
\caption{\label{vertex} Lepton number and CP-violation decays of $\wt
N_i$ with gaugino running in the vertex correction loop. Self-energy
contributions are ignored since they are suppressed by the extra
Yukawa couplings $Y^{N_i}$.}
\end{figure}

In our model, the dominant contributions to the lepton number production
come from the interference between  the  tree-level decays of $\wt N_i$
and the vertex corrections with gaugino running in the loop, which are
given in Fig.~\ref{vertex}. The original soft leptogensis relies
on the self-energy corrections to $\wt N_i$, and it is the  small mass
splitting (controlled by the bilinear soft terms
$B_{N_i}M_{N_i}\wt N_i^2$)
between the two real degrees of freedom of $\wt N_i$ denoted with $\wt
N_{i\pm}$ that resonantly enhances their CP-violation decays, see
Fig.~\ref{vertex}. To make the resonant effect  large enough,
 $B_{N_i}\lesssim 10^{-3} M_{SUSY}$ must be fine-tuned to be very
small~\cite{Grossman:2003jv,D'Ambrosio:2003wy}. Later,
 it was found that  the vertex corrections to $\wt N_i$ decays
with gaugino running in the loop contribute to the lepton number
asymmetry in a very different way, and then the normal value of
$B_{N_i}$ is allowed~\cite{Grossman:2004dz}. This is important for our model
to have successful soft leptogensis because its  UV completion
does not suppress  $B_{N_i}$.

In the previous Section, we have considered  the $N_2$ dominant
seesaw mechanism to produce $M_{LL}$. Corresponding to it,
this dominance again dominantly generate the lepton  asymmetry.
This can be seen clearly from
the  explicit calculations of the lepton  asymmetry produced by a
single $\wt N_i$ decays to lepton flavor $\alpha$, using the
procedure provided in  Ref.~\cite{Davidson:2008bu}
\begin{align}\label{asy}
 \epsilon_{i,\alpha}\equiv&
{\gamma( \wt L_\alpha H_u)+\gamma( L_\alpha \wt H_u)-\gamma( \wt
L_\alpha^\dagger H_u^\dagger)-\gamma( L_\alpha^\dagger \wt
H_u^\dagger)\over \sum_\beta \gamma( \wt L_\beta H_u)+\gamma(
L_\beta \wt H_u)+\gamma( \wt L_\beta^\dagger H_u^\dagger)+\gamma(
L_\beta^\dagger \wt H_u^\dagger)}\cr \approx& {3\alpha_2|Y^N_{\alpha
i}|^2\over4 \sum _{\beta}|Y^N_{\beta i}|^2}{M_2\over M_{N_i}}
\log{M_2^2\over M_2^2+M^2_{N_i}}\left(-{|A_{i2}|\over
M_{N_i}}\sin\theta_{A_{i2}}\right) \Delta_{BF}(T),
  \end{align}
where   $\alpha_2=g_2^2/4\pi$ and $g_2$ is $SU(2)_L$ gauge coupling.
This shows  the $N_2$ dominant contributions
from Eqs.~(\ref{LR}) and (\ref{nmass}), while the others are sub-dominant.
Note that $Y^{N_i}\sim 10^{-5}$ in our model,  the contributions from
self-energy corrections suppressed by $|Y^{N_i}|^2/\alpha_2$ are completely
ignorable. $\Delta_{BF}(T)$, whose expression is given in
 Ref.~\cite{Grossman:2003jv}, denotes for the thermal effect in the
thermal average of decay rates $\gamma$. Without it the above
asymmetry vanishes at zero temperature field theory due to the
exact cancellations between the fermionic and bosonic decay channels of
$\wt N_i$. By the way, our results are consistent with  a
previous work in Ref.~\cite{Fong:2009iu} which used
a different calculation method.

\begin{figure}[htb]
\begin{center}
\includegraphics[width=3.0in]{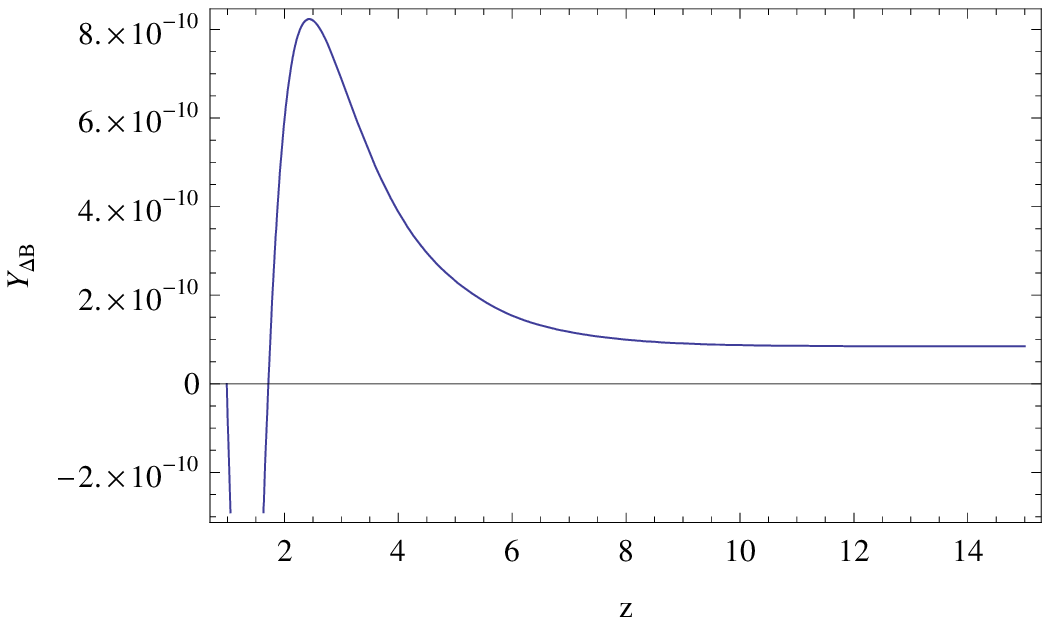}
\includegraphics[width=3.0in]{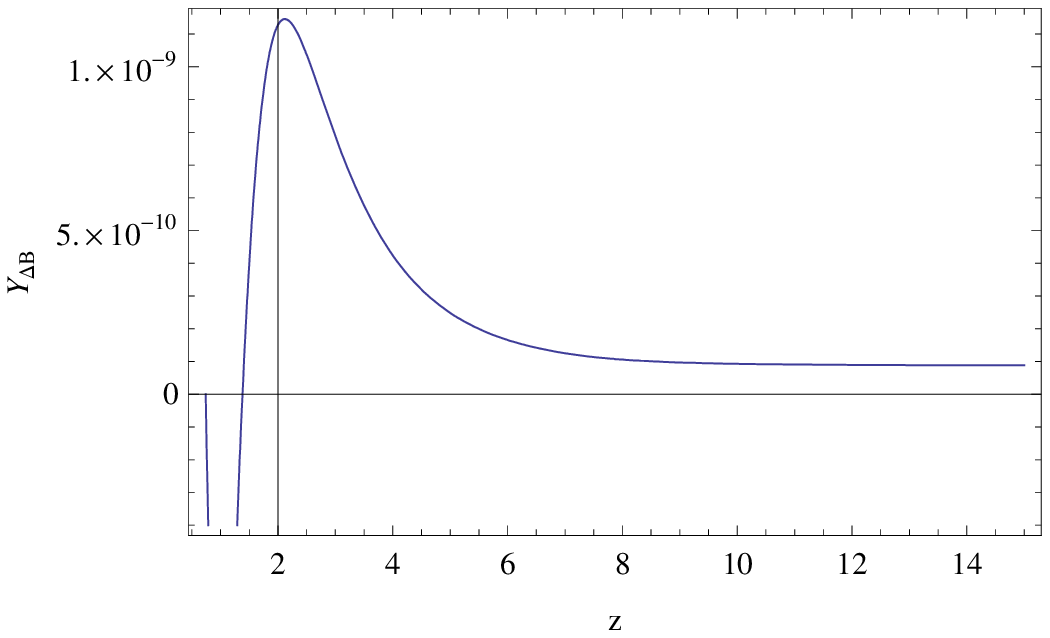}
\end{center}
\caption{\label{baryon} Baryon asymmetry $Y_{\Delta B}(z)$ versus
 $z=M_{N_2}/T$. Left: $M_{N_2}=10^{4}$ GeV, $M_{2}=250$ GeV and
$|A_{N_{22}}|=300$ GeV with $\theta_{A_{22}}=-1/4$. Right:
$M_{N_2}=4\times10^{3}$ GeV, $M_{2}=250$ GeV, $|A_{N_{22}}|=100$
GeV, $\theta_{A_{22}}=-1/5$. Initial density of $\wt N_2$ is taken
as thermal density.}
\end{figure}

The evolution of $\wt N_i$ lepton  number violation  decay and the
evolution of $\alpha$ ($\alpha=e,\mu,\,\tau$) lepton flavor number
are described by Boltzmann equations (BEs). Since the pure MSSM
interaction conserves the charge ${\Delta_\alpha}\equiv (B_f+B_s)/3-
(L_f+L_s)_\alpha$, where $(L_f)_\alpha=\ell_\alpha+e_\alpha$ and
$L_s$ is the total lepton number in scalar leptons. Similar
definition applies to the $B_{f,s}$. It is convenient to study the
evolution of density $\Delta_\alpha$, and the coupled BEs
are~\cite{Antusch:2006cw}
\begin{align}\label{BES}
\Delta_{\wt N}'=&-\sum_\alpha S_\alpha(z)-\left(Y_{\wt
N}^{eq}\right)',\quad S_\alpha(z)={z\over Y_{\wt N}^{eq}
}{\gamma_\alpha(z)\over
sH_N}\Delta_{\wt N},\\
 Y_{\Delta_ \alpha}'
=&-\epsilon_\alpha(z) S_\alpha(z)+W_{\alpha}(z) \sum_\beta
\left(A_{\alpha\beta}+C_\beta\right)Y_{\Delta_\beta},
\end{align}
where the derivative is on  $z\equiv M_{\wt N}/T$,  and  $\Delta_{\wt
N}\equiv Y_{\wt N}-Y_{\wt N}^{eq}$. In this crude set of BEs,
we consider the $\Delta L=1$  two to two
 scattering processes that provide the  CP asymmetry source, as
well as the wash-out from the top quark and gauge boson interactions.
 Also, the flavor effect is kept for the low-scale soft
leptogensis via  the $A$ matrix which expresses
$Y_{\ell_\alpha}$ as the linear combination of
$Y_{\Delta_\alpha}$~\cite{Barbieri:1999ma}, and via   the  $C$ matrix
which relates $Y_{{H_u}}$ to $Y_{\Delta_\alpha}$~\cite{Nardi:2006fx}
\begin{align}
A={1\over 207}\times\left(
\begin{array}{ccc}
 -64 & 5 & 5 \\
 5 & -64 & 5 \\
 5 & 5 & -64
\end{array}
\right),\quad C_\beta = {1\over 9} \sum_\alpha
A_{\alpha\beta}=-{2\over 69}\left(
                                                                 \begin{array}{c}
                                                                   1 \\
                                                                  1 \\
                                                                   1 \\
                                                                 \end{array}
                                                               \right).
\end{align}
 Our matrix is different from that in Refs.~\cite{Antusch:2006cw,Fong:2009iu}
since in this model the soft leptogenesis
proceeds during the era  $T\sim M_N\simeq 10^{4}{\rm \, GeV}$, where
all the Yukawa couplings and   the CKM mixings are in the chemical
equilibrium. Also, the $C$ matrix is included  here since its
entries are larger than the mixing entries in $A$ matrix.  Moreover, in
the simplified BEs, the source term and wash-out terms can be
rewritten analytically as follows
\begin{align}
 S_\alpha(z)&=z{K_1(z)\over K_2(z)}\mathcal
{K}_\alpha ,\quad
 W_\alpha(z)={1 \over 4}z^3{K_1(z)}\mathcal
{K}_\alpha ,
\end{align}
where the  object $\mathcal {K}_\alpha $ describes  the degree of
washout for a single flavor $\alpha$
\begin{align}
 \mathcal {K}_\alpha&\equiv{\Gamma_\alpha+\wt \Gamma_{\alpha}\over
H_N}={m_\alpha\over m_{MSSM}^*},\cr
m_\alpha&=|Y^N_{\alpha2}|^2v^2\sin^2\beta/M_{N_2},
\end{align}
where $m_{\alpha}$ is equal to ${|B|^2/ M_{N_2}}$ for ${\alpha=2,3}$
while vanishes for $\alpha=1$. In the  MSSM using $g_*=228.75$ at
$T\sim M_N$, one obtains $m_{MSSM}^*\approx \sin^2\beta\times 1.58\times
10^{-3}$eV.  Thus, $\mathcal {K}_{2,3}\sim 20$, and then the
soft  leptogensis is in the  strong wash out region~\cite{Barbieri:1999ma,
Davidson:2008bu}. Finally, the sphaleron processes transform the
survival lepton asymmetry into baryon asymmetry, eventually gives
the baryon number density
\begin{align}
Y_{B}^{MSSM}={n_B-n_{\bar B}\over s}= {10\over
31}\sum_{\alpha}Y_{\Delta_{\alpha}}~.~\,
  \end{align}
With the initial density of $\wt N_i$ in thermal
equilibrium required by a successful  freeze-in mechanism,
we present the numerical solutions to the baryon asymmetry evolution in
Fig.~\ref{baryon}.

The observed baryon asymmetry
$Y_{B}=(8.75\pm0.23)\times 10^{-11}$~\cite{Komatsu:2008hk}
is generated with the following  parameters:
$M_{N_2}=10^{4}$ GeV, $M_{2}=250$ GeV and $|A_{N_{22}}|=300$ GeV
with phase $\theta_{A_{22}}=-1/4$.  As the RHN mass decreases,
for instance,
$M_{N_2}=4\times10^{3}$ GeV, the baryon asymmetry tends to be
overproduced. The reason is that the lepton asymmetry given in Eq.~(\ref{asy}) is
proportional to $1/M_{N_2}^2$ but linear to $A_{N_2}$ and $M_2$.
Because we have assumed that the LSP $\wt G$ has mass about $ 200$ GeV,
$M_2$ can not be too small. Therefore, we can choose a  smaller
$|A_{N_{22}}|=100$ GeV with phase $\theta_{A_{22}}=-1/5$,
or we can fine-tune the phase of $A_{N_{22}}$. Anyway,
the observed baryon asymmetry can be obtained  in the general parameter
space.

\section{ Discussions and Conclusion }

Cosmic ray anomalies from the Fermi-LAT and PAMELA
experiments can be naturally explained by the
TeV-scale decaying DM with a very long
lifetime $\sim 10^{26}$s which decays dominantly
to the muon and tau leptons.
Note that the neutrino TBM can be realized elegantly via
the $\mu-\tau$ symmetry, we conjectured that
the DM decay is related to the neutrino physics.
We considered the supersymmetric Standard Model
with three right-handed neutrinos. To realize
the decaying DM, we introduced a $Z_3$ discrete
symmetry and two DM particles $\phi_1$ and $\phi_2$.
Because $\phi_1 \phi_2$ can couple to the right-handed
neutrinos via the dimension-five operators suppressed by
the GUT scale $ M_{GUT}$, DM particle has a natural lifetime
around  $\tau\sim 10^{26}s$ if the
seesaw scale is about $10^4$ GeV.
In particular, the DM particle  will decay dominantly
to the $\mu$ and $\tau$ final states due to the
 $N_2$ dominant seesaw mechanism.
 Moreover, the DM relic density, which usually is a
problem in decaying DM models, can be
  achieved naturally through the freeze-in mechanism
  with couplings typically about $ {\cal O}({\rm TeV})/ M_{GUT}$.
 Simultaneously the small scale problem on the power spectrum
can be solved since  the
 metastable particles in the DM sector, which are also freezed into
the thermal bath, can decay to the relativistic LSP in the
supersymmetric SMs.
Furthermore, we showed that the baryon asymmetry can be generated
via the  soft leptogenesis in a large region of the
parameter space for supersymmetry breaking soft mass terms.

\begin{acknowledgments}

We would like to thank S. Matsumoto for helpful discussions.
This research was supported in part
by  the DOE grant DE-FG03-95-Er-40917 (TL),
by the Natural Science Foundation of China
under grant No. 10821504,
and by the Mitchell-Heep Chair in High Energy Physics (TL).

\end{acknowledgments}

%%%%%%%%%%%%%%%%%%%%%%%%%%%%%%%%%%%%%%%%%%%%%%%%%%%%

%%%%%%%%%%%%%%%%%%%%%%%%%%%%%%%%%%%%%%%%%%%%%%%%%%%%%%%%%%%%%%%%%%%%%%%%%%%%%%%%

%%%%%%%%%%%%%%%%%%%%%%%%%%%%%%%%%%%%%%%%%%%%%%%%%%%%%%%%%%%%%%%%%%%%%%%%%%%%%%%%

\appendix

\section {A Brief Review of  Freeze-In Mechanism }
\label{freezein}

In this Appendix we will give a brief review of the freeze-in
mechanism, but we shall formulate it differently.
The basic idea of freeze-in mechanism is that in the BE for FIMP $X$,
there is no inverse decay
or scattering process to the mother particle that produces FIMP,  as a
result even small interaction rate is also successful in generating
significant relic density for FIMP. As a starting  point, the simple
BE for $X$ is (We consider the scattering process as an example,
and  the  similar expression holds for decay.)
\begin{align}\label{Y}
Y'(z)={s\,z\over H_1}{\gamma(A+B\rightarrow X+C)\over( s Y^{eq}_A)(
s Y^{eq}_B)}Y_A(z)Y_B(z),
\end{align}
where the  Hubble constant at $T=M_A$ is
$H_1(T)|_{M_A}=1.66\sqrt{g_*^\rho}T^2/M_{pl}|_{M_A}$. In this paper,
we use ${g_*^{\rho,s}}$ to denote the effective numbers of degree of
freedom in the thermal bath at the freeze-in temperature  $T\sim
M_{A}$, respectively for the entropy density $s$ and energy density
$\rho$.

If $A$ and $B$ are assumed to be in thermal equilibrium during
freeze-in,  the BE reduces to the situation
 discussed in Ref.~\cite{Hall:2009bx}.
The FIMP is produced dominantly at the temperature around the mass of
heavier bath particles, when the bath particles still track their
equilibrium distribution closely~\cite{Hall:2009bx}.
So the approximations are valid. In
our paper,  from Eq.~(\ref{BES}) one finds that $\wt N$ and $N$
deviate from  their  equilibrium typically at $T_f\sim M_N/5$ (due to
strong washout, sRHNs departure from equilibrium rather late), so
the equilibrium approximation is also employed here.

 In our model, FIMP is freezed-in both from $ N_{\pm}$ decays and
 their scatterings. First let us discus the decay.  The yield of $X$  is produced
simply  by integrating the right-hand side of Eq.~(\ref{Y}) over  $z$
from $0$ to $\infty$. A good property of freeze-in mechanism is that
$Y(x)$ is insensitive to the UV physics, which is obvious from the
integrand. So one can safely ignore the time of lower bound which
may sensitive to inflation or reheating at the UV. For two-body
decay, the thermally averaged decay rate is easily obtained analytically
\begin{align}
\gamma(A\rightarrow X+C)= {g_AT^3\over2 \pi^2}z^2
K_1(z)\Gamma(A\rightarrow X+C),
\end{align}
with $g_A$ the internal degrees of freedom of $A$. Furthermore,
let us reasonably assume that the freeze-in process lasts from
$z\approx 0$ till  $z\gtrsim \mathcal {O}(1)$ when generally the
weakly interacting particle $A$ decouples from thermal bath. Then
one has
\begin{align}
Y(z\gtrsim 10)\approx&\int_0^{\infty}dz{z\over
sH_1}\gamma(A\rightarrow X+C)\cr =& {135 \,g_A\over 8\pi^3 (1.66)
g_*^s \sqrt{g_*^\rho}}{M_{pl}\Gamma(A\rightarrow X+C)\over M_A^2}.
\end{align}
Then the  relic density of $X$ is given by
\begin{align}\label{relic}
\Omega_Xh^2\approx 4.50\times 10^{25}\times \lambda^2{g_A\over
g_*^s\sqrt{g_*^\rho}}{M_X\over M_A},
\end{align}
where we  have typically used $\Gamma(A\rightarrow
X+C)=\lambda^2M_A/8\pi$ and dropped  the phase space factor. If
multi thermal particles $A_i$ contribute to freeze-in, $i$ should be
summed over. Next we study the freeze-in mechanism
through scattering processes.
The thermally averaged scattering rate is formally given by
\begin{align}
\gamma(A+B\rightarrow X+C) &= {g_Ag_B T^6\over
16\pi^4}\int_{(m_A+m_B)^2/s}^\infty dx \,x^4
K_1(x)\lambda(1,x_A,x_B)
\nonumber \\
& \times \sigma(A+B\rightarrow X+C),
\end{align}
where $x=\sqrt{s}/T$, and then the integrand depends on $z$
through its $s$ dependence.
 Similarly, the final yield of $X$
is obtained by integrating over $z$ from 0 to some large value
\begin{align}
Y_X(\infty)\approx&\int_0^{\infty}dz{z\over
sH_1}\gamma(A+B\rightarrow X+C)\cr =& {45\, g_Ag_B \over 2\times
1.66\times 256\pi^7 g_{*}^s\sqrt{g_{*}^\rho }}{M_{pl}\over
M_A}\times \mathcal {I}[x,z],\cr
 \mathcal {I}[x,z]\equiv& \int_0^{\infty} dz \int_z^{\infty}dx\, x^2
 K_1(x) \Xi(x,z)\sim \mathcal {O}(1),
\end{align}
where  $ \Xi(x,z)=(16\pi s)\lambda(1,x_A,x_B)\sigma(x,z)$, and
$\sigma(x,z)$ is the scattering  cross section. Eventually, the
relic density is
\begin{align}\label{22S}
\Omega_Xh^2\approx&6.0\times 10^{22}\times{ g_Ag_B \over
g_{*}^s\sqrt{g_{*}^\rho }}\times\left({M_X\over
M_A}\right)\times\mathcal {I}[x,z].
\end{align}

%%%%%%%%%%%%%%%%%%%%%%%%%%%%%%%%%%%%%%%%%%%%%%%%%%%%%%%%%%%%%%%%%%%%%%%%%%%%%%%%

%%%%%%%%%%%%%%%%%%%%%%%%%%%%%%%%%%%%%%%%%%%%%%%%%%%%%%%%%%%%%%%%%%%%%%%%%%%%%%%%

%%%%%%%%%%%%%%%%%%%%%%%%%%%%%%%%%%%%%%%%%%%%%%%%%%%%%%%%%%%%%%%%%%%%%%%%%%%%%%%%

%%%%%%%%%%%%%%%%%%%%%%%%%%%%%%%%%%%%%%%%%%%%%%%%%%%%%%%%%%%%%%%%%%%%%%%%%%%%%%%%

\end{document}